\documentclass[aps,prl,twocolumn,groupaddress,showpacs]{revtex4-1}
\usepackage{amsmath,amssymb,graphicx,stmaryrd}

\usepackage{ifthen}
\newboolean{PRLVERSION}
\setboolean{PRLVERSION}{false}

\ifthenelse {\boolean{PRLVERSION}}
{
\newcommand{\Letter}{Letter}
\newcommand{\co}{(color online)\ }

\newcommand{\centred}{centered}

\newcommand{\favour}{favor}
\newcommand{\Acknowledgements}{Acknowledgments}
\newcommand{\doivpy}[2]{#2}
\newcommand{\arxiv}[2]{arxiv:#1 #2}
\newcommand{\isbn}[1]{}
}
{
\pdfoutput=1
\usepackage{color}
\definecolor{LinkColor}{rgb}{0.256,0.439,0.588}
\usepackage{hyperref}
\hypersetup{
   pdfauthor={KSD Beach, Jin Xu},
   pdftitle={Two distinct spin liquid states in a layered cubic lattice},
   pdfsubject={Quantum spin liquids},
   pdfkeywords={resonating valence bonds,} {cubic lattice,} {anisotropy,} {spin liquid},
   colorlinks=true,
   citecolor=LinkColor,
   linkcolor=LinkColor,
   urlcolor=LinkColor
   }
\usepackage{mathptmx}
\newcommand{\Letter}{paper}
\newcommand{\co}{}

\newcommand{\centred}{centred}

\newcommand{\favour}{favour}
\newcommand{\Acknowledgements}{Acknowledgements}
\newcommand{\doivpy}[2]{\href{http://dx.doi.org/#1}{#2}}
\newcommand{\arxiv}[2]{\href{http://arxiv.org/abs/#1}{arxiv:#1 #2}}
\newcommand{\isbn}[1]{ISBN #1}
}

\usepackage{bm}
\usepackage{bbold}

\newcommand{\ket}[1]{\lvert #1 \rangle}
\newcommand{\expectation}[1]{\langle #1\rangle}
\newcommand{\overlap}[2]{\langle #1 | #2 \rangle}
\newcommand{\matrixelem}[3]{\langle #1 | #2 | #3 \rangle}

\begin{document}

\title{Two distinct spin liquid states in a layered cubic lattice}

\author{Jin Xu}
\author{K. S. D. Beach} \email{kbeach@ualberta.ca}
\affiliation{Department of Physics, University of Alberta, Edmonton, Alberta, Canada T6G 2E1}

\date{October 31, 2013}

\begin{abstract}
We construct a family of short-range resonating-valence-bond wave functions on a layered cubic lattice, 
allowing for a tunable anisotropy in the amplitudes assigned to nearest-neighbour valence bonds along 
one axis. Monte Carlo simulations reveal that four phases are stabilized over the full range of the anisotropy 
parameter. They are separated from one another by a sequence of continuous quantum phase transitions.
An antiferromagnetic phase, {\centred} on the perfect isotropy point, intervenes between two {\it distinct} quantum
spin liquid states. One of them is continuously deformable to the two-dimensional U(1) spin liquid, 
which is known to exhibit critical bond correlations. The other has both spin and bond correlations 
that decay exponentially. The existence of this second phase is proof that, contrary to expectations,
neither a bipartite lattice structure nor a conventional Marshall sign rule is an impediment to realizing a 
fully gapped quantum spin liquid.
\end{abstract}

\maketitle

A quantum spin liquid~\cite{Balents2010} is an exotic insulator that breaks no 
symmetries down to zero temperature. The picture is that of a Mott phase~\cite{Mott49} with quenched charge fluctuations,
whose residual degrees of freedom are governed by a low-energy model of lattice spins coupled
through local interactions. By some confluence of disruptive factors---quantum fluctuations, un{\favour}able geometry, 
exchange interactions that are at odds with one another---the system is unable to establish any kind of long-range order.
Over the years, a large number of gapped and ungapped
liquid states has been proposed~\cite{Baskaran87,Anderson87,Affleck88a,Affleck88b,Dagotto88,Kotliar88,Read89a,Read91,Wen91,Lee96},
and there is sufficient evidence to believe that such states are found in real materials~\cite{Shimizu03,Pratt11,Han12}.

There has been a spate of recent numerical results supporting the existence of gapped spin liquid
ground states in simple frustrated Heisenberg models~\cite{Jiang08,Yan11,Depenbrock12,Jiang12a,Wang13}.
These are believed to be related to states with $\mathbb{Z}_2$ topological 
order~\cite{Read91,Wen91,Wegner71,Senthil00,Moessner01a,Moessner01b}.
There are, however, serious practical difficulties in (convincingly) connecting spin models to their purported 
spin liquid ground states. In particular, obtaining unbiased, well-converged results is
a challenge. Quantum Monte Carlo simulation is hindered by the infamous sign 
problem~\cite{Troyer2005}; and beyond one spatial dimension (1D), the computational 
complexity for density-matrix renormalization group calculations scales exponentially 
in the lattice size~\cite{Schollwock2005}. So there remains some question about the 
reliability of numerical results and ongoing controversy as to whether some of these 
delicate liquid states might actually be unstable to valence bond 
crystal order~\cite{Read89b,Kotov99,Singh08,Sandvik12}.

A somewhat less fraught path is to dispense with the microscopic model entirely and simply to construct
liquid states for study. Considerable analytical work has been done (using slave-particle approaches, 
gauge field theories, and the ideas of projective symmetries and cohomology groups)
to develop classication schemes~\cite{Wen02,Wang06,Tchernyshyov06,Essin13}.
On the numerical side, there has been an active effort to 
construct and characterize trial wave functions that are designed to be featureless.
These calculations have generally been carried out in the context
of resonating-valence-bond (RVB) states~\cite{Pauling49,Anderson73}, which are total-spin-zero
states built out of pairs of spins forming singlets. In a few cases, and with some success, the RVB 
states have been treated variationally~\cite{Lou07,Li12,Hu13,Zhang13a,Zhang13b}.
But more often, the program is simply to consider 
particular RVB wave functions, either evaluated directly~\cite{Albuquerque10,Tang11,Albuquerque12,Ju12};
recast as Pfaffians~\cite{Wildeboer12}; or Gutzwiller-projected from a free Fermi sea~\cite{Zhang11} 
or a Bardeen-Cooper-Schrieffer (BCS) state~\cite{Yang12}.

This is akin to the approach we take in this {\Letter}, except that we also allow for a modulation of the spatial 
anisotropy. We construct a {\it family} of short-range RVB states
on the cubic lattice with a single control parameter that represents the relative probability for a valence bond
to be oriented along one special axis. We find that the phase diagram includes two disordered
quantum states---one of which is surprising in that all its correlations decay faster than a power law,
implying a gapped quantum spin liquid.

What animates our study is the conjecture by Yang and Yao that spin and bond correlations decay
exponentially in all short-range, nonbipartite RVB states~\cite{Yang12}. (This is suggested by the 
special role that a nonbipartite lattice plays in achieving the $\mathbb{Z}_2$ 
topological state in quantum dimer models~\cite{Moessner01a,Moessner01b}.) What has been
unclear until now is whether the converse also holds. In other words, is frustration at the level of 
the wavefunction a {\it necessary condition} for a gapped spin liquid? As it turns out, no---but there has
been good reason to think it might.
It is certainly true that on the square lattice the NN RVB state is understood to be a U(1) spin liquid with 
critical bond correlations~\cite{Albuquerque10,Tang11}; and the same state transplanted to
the cubic and diamond lattices exhibits bond correlations with a dipolar form~\cite{Albuquerque12} 
characteristic of the Coulomb phase~\cite{Henley10}. Indeed, in all the currently studied examples of 
a short-range RVB state on a bipartite lattice, there are remnants of critical correlations inherited 
from the underlying set of classical hardcore dimer tilings.
The results reported here, however, provide a clear counterexample.

\begin{figure}
\begin{center}
\ifthenelse {\boolean{PRLVERSION}}
{\includegraphics[scale=0.9]{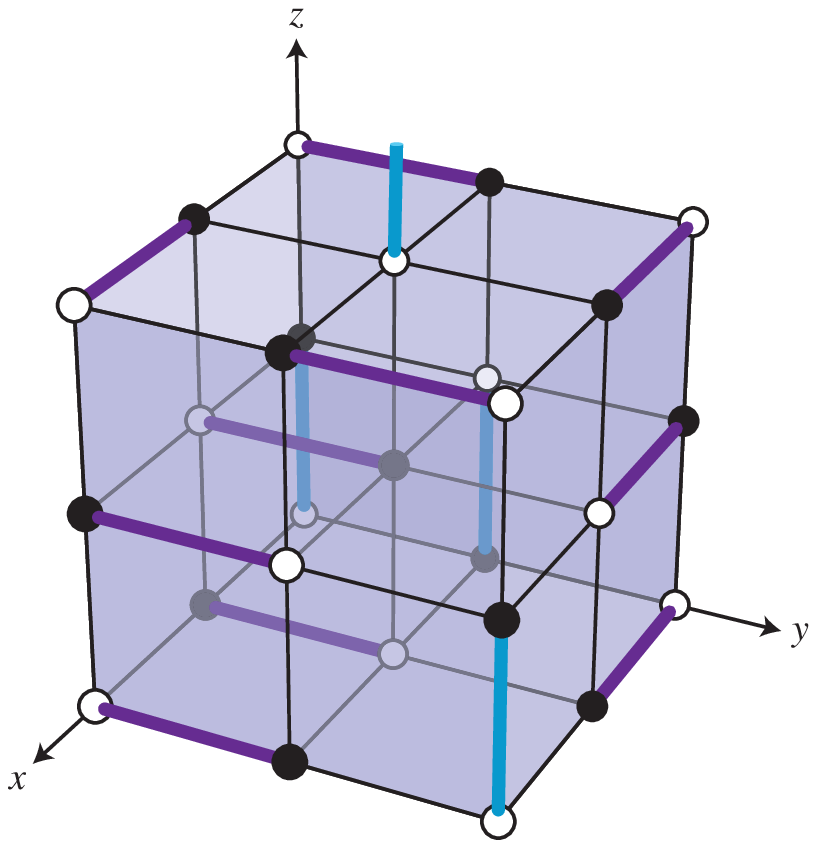}}
{\includegraphics[scale=0.9]{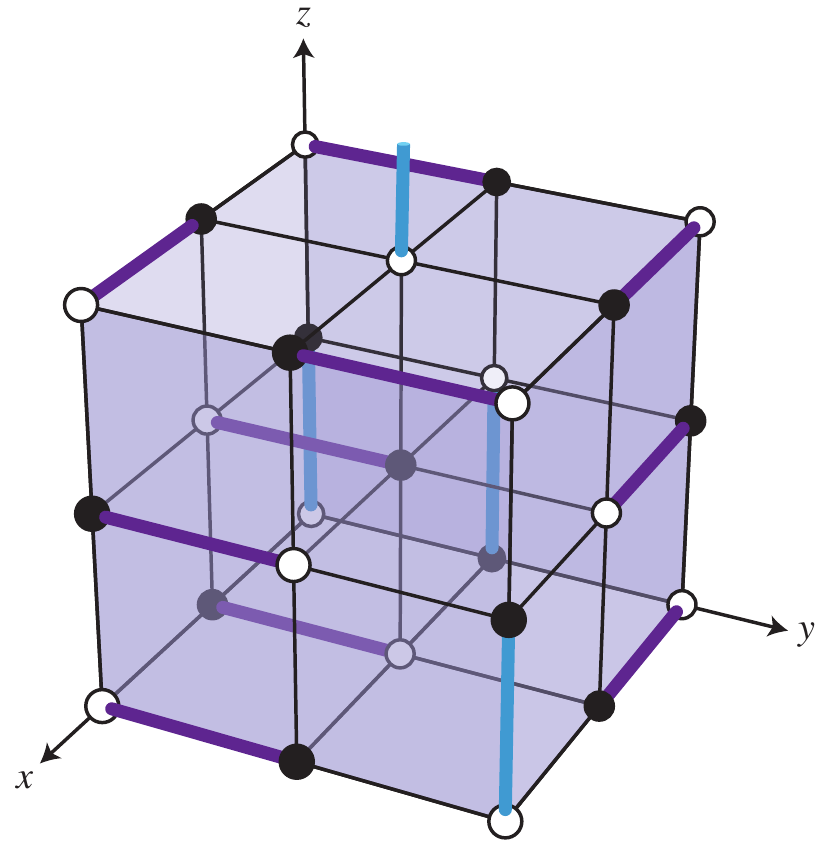}}
\end{center}
\vskip-4mm
\caption{\label{FIG:model}\co 
A snapshot of the resonating short-bond-only state on a layered cubic lattice. Valence bonds lying in the $xy$ 
plane (dark purple) are distinguished from those aligned with the $z$ axis (light blue).
Each $z$-oriented bond contributes an additional factor $a$, which encodes its probability
relative to bonds of the other two orientations.
}
\end{figure}

{\it Model}.---We consider a system of $S = 1/2$ spins, $2N = L^3$ in number, living on a cubic lattice of linear size $L$
with periodic boundary conditions. The wave function is taken to be a superposition of valence bond states, and we treat their
coefficients (in the style of Liang, Doucot, and Anderson~\cite{Liang88}) as a product of
individual bond amplitudes
\begin{equation} \label{EQ:h_def}
h(x,y,z) = \delta_{|x|+|y|+|z|,1}\bigl[ 1 + (a-1)\delta_{|z|,1} \bigr].
\end{equation}
Equation~\eqref{EQ:h_def} encodes a very simple rule: the amplitude is 1 for any nearest-neighbour (NN) bond 
oriented along the $x$ or $y$ axes, $a > 0$ for any NN bond oriented along the $z$ axis, and zero otherwise.

Hence, as illustrated in Fig.~\ref{FIG:model}, the wave function realizes a one-parameter family of short-range RVB 
states living on an effectively layered cubic lattice, with quite different easy-plane ($a \ll 1$) and easy-axis ($a \gg 1$)
limits. 
The wave function can be expressed concisely as
\begin{equation}
\ket{\psi} = {\sum_v}' a^{\,n_z(v)} \ket{v},
\end{equation}
where $\ket{v}$ is a valence bond state; the prime indicates 
a sum over all bipartite, short-bond-only configurations;
and $n_z(v)$ counts the number of bonds parallel to the $z$ axis.

Including only NN valence bonds is a reasonable choice. Such a restriction emerges naturally 
in some exact solutions~\cite{Fujimoto05,Seidel09,Cano10}, and it is has been judged an excellent
approximation in the case of many strongly frustrated antiferromagnets~\cite{Mambrini06,Flocke98,Nussinov07,Mosadeq11}. 
A crucial observation is that when long bonds are suppressed in two-dimensional (2D) systems, the accompanying N\'eel 
order is suppressed too, but this is not always the case in 3D.
Specifically, the equal-amplitude NN RVB wave function on the square lattice turns out to be 
disordered~\cite{Albuquerque10, Tang11}, whereas the same state on the cubic
lattice has a substantial staggered moment 
$m_\text{s} = 0.1519(5)$~\cite{Beach07b,Albuquerque12}. 

Our trial wave function on a layered cubic lattice is designed to interpolate between these two limits. 
Therefore, it must be that, somewhere between the 2D spin liquid ($a=0$) and the isotropic 3D N\'eel 
state ($a=1$), there is a phase transition at some critical value $a_1$.  Moreover, we are free to tune the anisotropy in the 
opposite direction. In the limit $a\to\infty$, the system consists of $L^2$ decoupled chains, 
each of them a perfectly ordered 1D bond crystal (as occurs in the Majumdar-Ghosh 
chain~\cite{Majumdar69}). Hence, we anticipate an additional pair of critical points, 
$a_2$ and $a_3$, at which the N\'eel order is extinguished and the bond order emerges.
What is so exciting is that these two points turn out not to be coincident, and an additional disordered 
phase occupies the region $a_2 < a < a_3$.

{\it Numerical results}.---The wave function is evaluated using Monte Carlo
with worm-like updates~\cite{Zhang13a}.
Expectation values of observables 
\begin{equation}
\expectation{\hat{O}} = \frac{1}{Z} {\sum_{c}}' w(c) O(c)
\end{equation}
are sampled with respect to the probability distribution $w(c)/Z$, whose domain
is the set of all (closed-loop-forming)
short-range double bond coverings $c = (v_1, v_2)$. Here, 
the factor $Z = \overlap{\psi}{\psi} = \sum_{c}' w(c)$ fixes the overall normalization. 
The weight $w(c) = w(v_1,v_2) = a^{n_z(v_1) + n_z(v_2)} \overlap{v_1}{v_2}$
is guaranteed to be positive definite, provided that $a > 0$; since the basis of 
nearest-neighbour bond states is bipartite on the cubic lattice, 
the overlap $\overlap{v_1}{v_2}$ is positive definite~\cite{Beach08} and is given by a simple power
of 2 as dictated by the transition graph~\cite{Sutherland88}.
Operator expectation values $O(c) = {\matrixelem{v_1}{\hat{O}}{v_2}} / \overlap{v_1}{v_2}$ are computed according
to the loop estimators given in Ref.~\onlinecite{Beach06}. 

In order to track the magnetism, 
we measure the spin correlation function $C_{\text{s}}(x,y,z) = \expectation{\mathbf{S}(0,0,0) \cdot \mathbf{S}(x,y,z)}$, 
the square of the $(\pi,\pi,\pi)$ staggered moment
\begin{equation}
\langle \hat{m}_\text{s}^2 \rangle
= \frac{1}{L^3} \sum_{x,y,z=1}^L C(x,y,z) (-1)^{(x+y+z)},
\end{equation}
and $U = 1-3\expectation{\hat{m}_\text{s}^4}/5\expectation{\hat{m}_\text{s}^2}^2$, the corresponding 
Binder cumulant.
Within our simulations, the appearance of long-range antiferromagnetic correlations is signalled by 
the proliferation of system-spanning loops. In terms of our update algorithm, we expect N\'eel ordering 
to coincide with the worm head and tail (i.e., the end-points of the evolving open string that serves as 
the Monte Carlo walker) becoming deconfined and thus free to circumnavigate the periodic lattice independently. 
With that in mind, we also compute $\expectation{W^2} = \expectation{W_x^2}+\expectation{W_y^2}+\expectation{W_z^2}$,
the average of the squared winding number of the worm path,
summed over each of the three orthogonal directions.

\begin{figure}
\begin{center}
\ifthenelse {\boolean{PRLVERSION}}
{\includegraphics{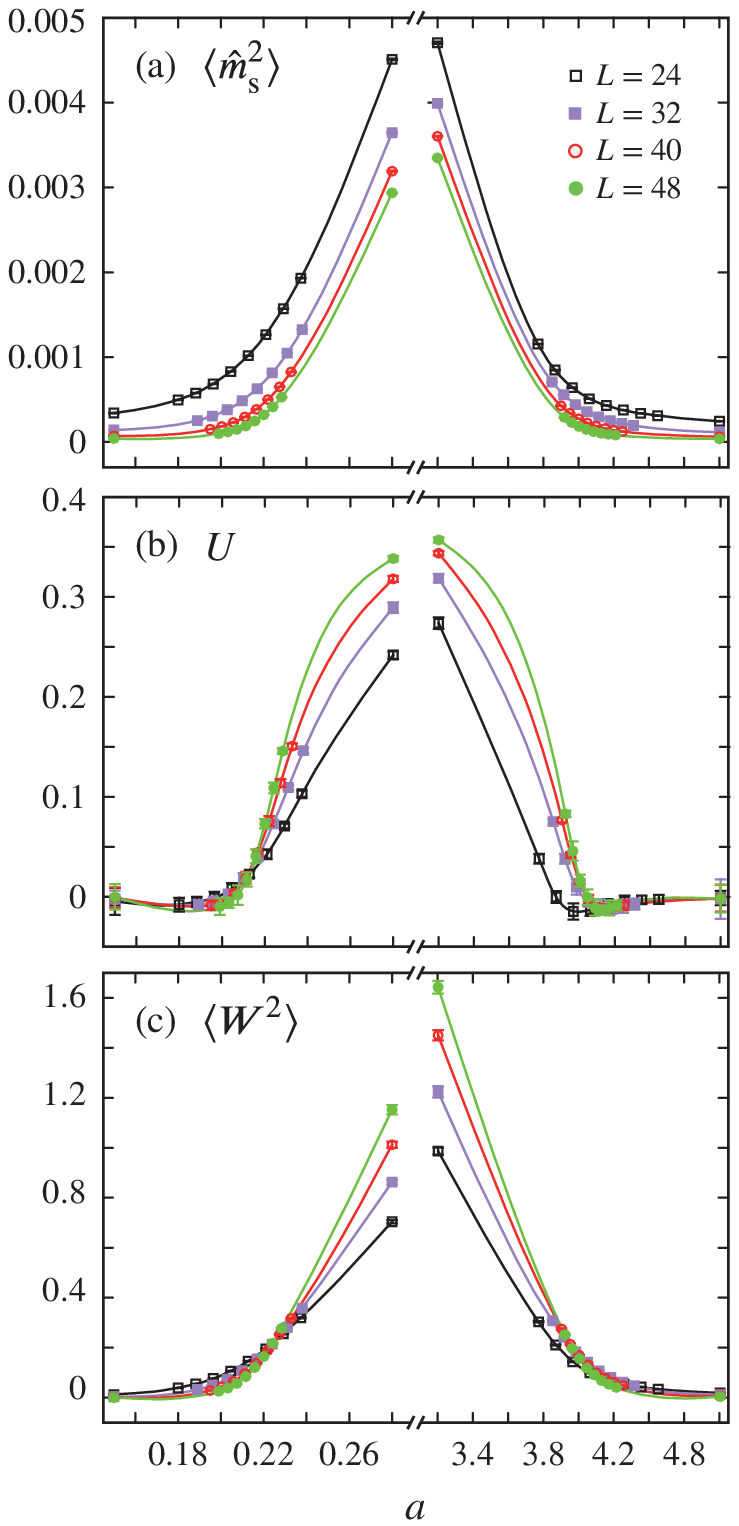}}
{\includegraphics{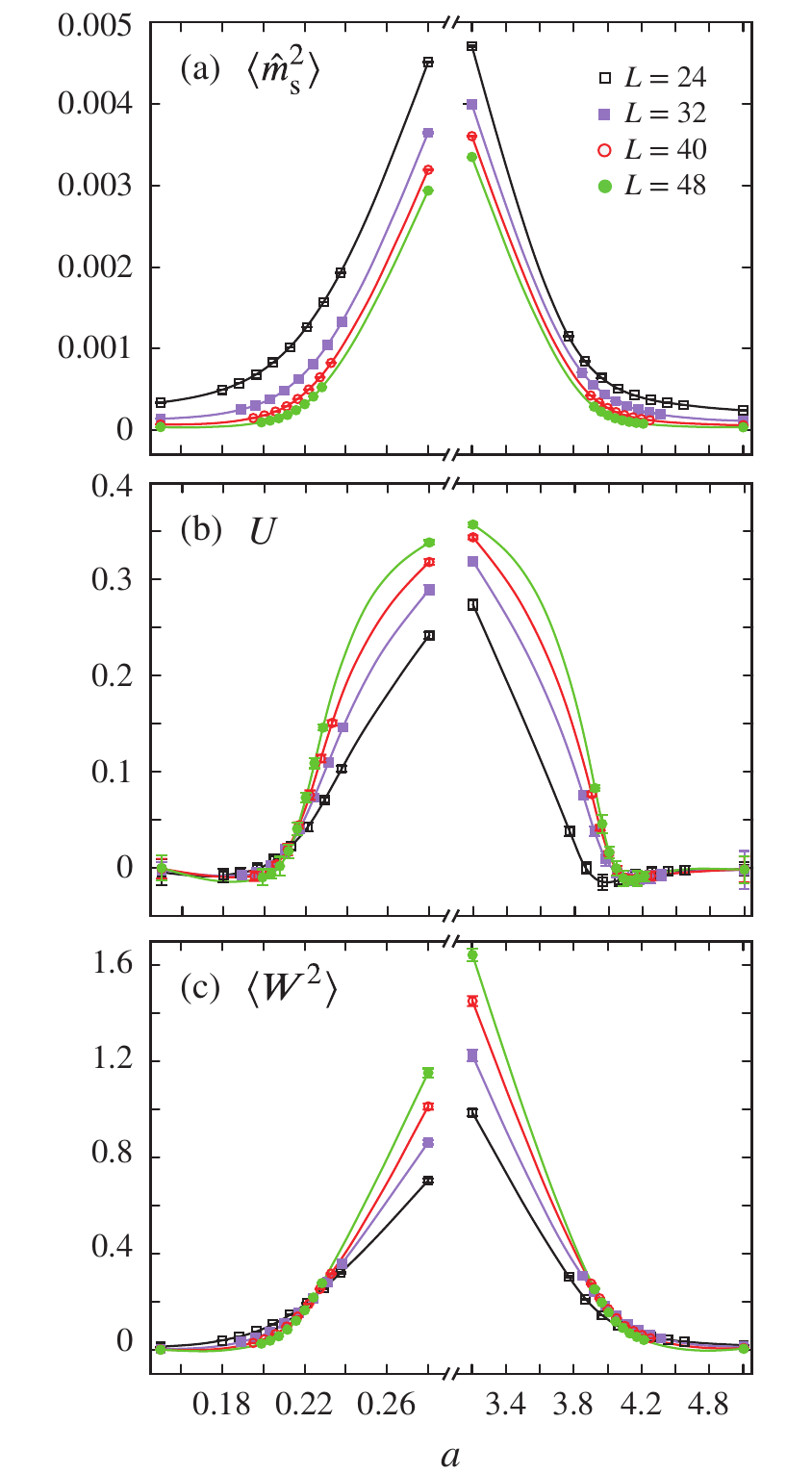}}
\end{center}
\vskip-4mm
\caption{\label{FIG:order_params}\co 
Various magnetization order parameters as a function of the anisotropy $a$ on cubic lattices of linear size $L = 24, 32, 40, 48$.
The quantities shown are (a) the square of the staggered magnetic moment, (b) the Binder cumulant, 
and (c) the worm winding order parameter.}
\end{figure}

As shown in Fig.~\ref{FIG:order_params}, we observe a continuous transition from the easy-plane
spin liquid to the N\'eel phase at $a_1 \approx 0.22$ and from the N\'eel phase to the 
easy-axis spin liquid at $a_2 \approx 3.9$.  
We estimate the critical points and exponents of the thermodynamic system from
finite-size simulations on the $L^3$ cubic lattice, performed for increasing values of the linear size up to $L=48$.
In our analysis, we assume the conventional, leading-order scaling form for the staggered magnetization:
\begin{equation}
\langle \hat{m}_{\text{s}}^{2k} \rangle = L^{-2k\beta/\nu}\mathcal{M}_{k,n}[(a-a_n)L^{1/\nu_n}].
\end{equation} 
Furthermore, we suppose that the Binder cumulant and (by analogy) our winding measurement scale according to
$U = \mathcal{U}_n[(a-a_n)L^{1/\nu_n}]$ and $\langle W^2 \rangle = \mathcal{W}_n[(a-\tilde{a}_n)L^{1/\mu_n}]$.
Here, $a_n$ is a stand-in for one of the critical anisotropies $a_1$ or $a_2$. The tilde marks the corresponding
value extracted from the winding data alone.

We make use of measurements from the largest three simulations sizes  ($L = 32, 40, 48$)
to determine the critical anisotropies and critical exponents. 
We bootstrap~\cite{Efron93} the underlying data and apply the fitting procedure to each resampled set
in order to establish the spread. The resulting values with error estimates are listed in 
Table~\ref{TAB:critical.values}.
Note that the critical anisotropies extracted (independently) from the magnetic and 
winding number data disagree slightly
for the first phase transition: $a_1$ and $\tilde{a}_1$ differ by about 4\%. Nonetheless, it is likely 
that $a_n = \tilde{a}_n$ for both $n=1$ and $n=2$ and that our analysis simply 
underestimates the systematic error that accrues from neglecting subleading corrections to 
the scaling form. 

The correlation length exponents at the opposite edges of the N\'eel-ordered region, $\nu_1$ and
$\nu_2$, are consistent with one another, 
but the large difference between the magnetization exponents,
$\beta_1$ and $\beta_2$, suggests that the two transitions are in different universality classes. A deconfinement transition is 
evident in the dynamics of the worm, with an (algorithm-dependent) 
exponent that appears 
to be exactly $\mu_1 = \mu_2 = 1$.

\begin{table}
\setlength{\tabcolsep}{5pt}
 \begin{tabular}{llllcll} \hline\hline
 & \multicolumn{3}{c}{magnetism} && \multicolumn{2}{c}{winding}\\ 
 \cline{2-4}\cline{6-7}
$n$ & $a_n$ & $\nu_n$ & $\beta_n$ &&$\tilde{a}_n$ & $\mu_n$\\ \hline
1 & 0.2157(4) & 0.768(3) & 0.85(1) && 0.2240(2) & 1.00(2)  \\
2 & 3.944(5) & 0.764(7) & 0.65(2) && 3.954(3) &0.98(2)  \\ \hline\hline
\end{tabular}
\caption{\label{TAB:critical.values}The critical points and exponents extrapolated from a finite-size
scaling analysis of the simulation data for $L = 32, 40, 48$.}
\end{table}

Finally, we also consider spatially resolved measurements of the spin correlations $C_\text{s}(x,y,z)$
and of the correlations
\begin{equation} \label{EQ:bond_corr}
C_\text{b}(x,y,z) = \expectation{P(0,0,0)P(x,y,z)} - \expectation{P(0,0,0)}\expectation{P(x,y,z)}
\end{equation}
between $z$-directed valence bonds, as detected by the singlet projection operator
$P(x,y,z) = 1/4-\mathbf{S}(x,y,z) \cdot \mathbf{S}(x,y,z+1)$. 
When $0 < a < a_1$, we think of the system as consisting of $L$ stacked copies of the U(1) spin liquid, 
weakly coupled. Indeed, we find that spin correlations in all directions are short-ranged and that
there are power law bond correlations confined to each $xy$ layer.
On the other hand, when $a_2 < a < a_3$, all correlations decay faster than a power law.

\begin{figure}
\begin{center}
\ifthenelse {\boolean{PRLVERSION}}
{\includegraphics{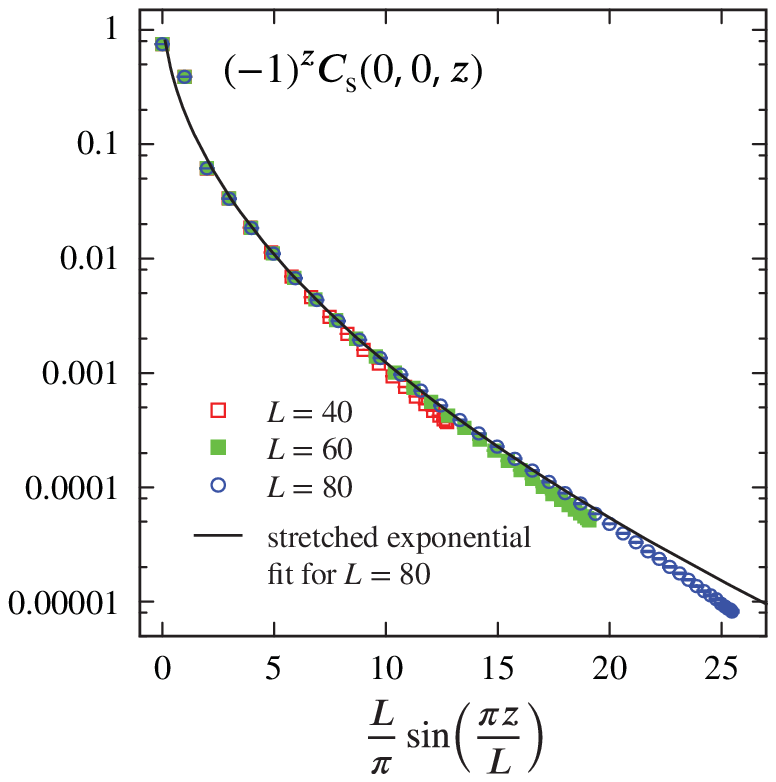}}
{\includegraphics{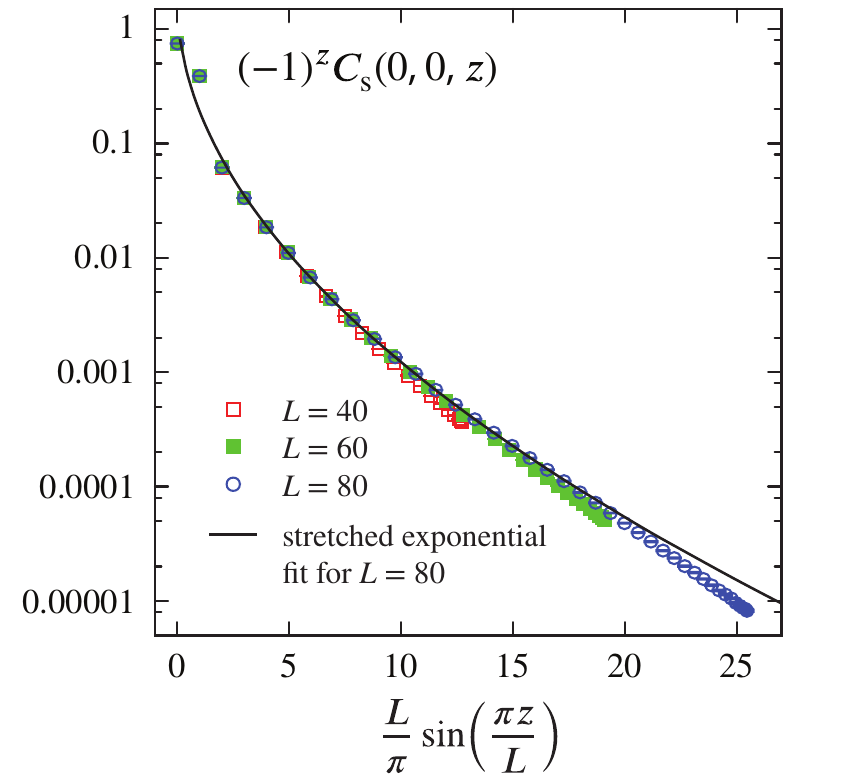}}
\end{center}
\vskip-4mm
\caption{\label{FIG:spin_corr}\co 
The magnitude of the alternating spin correlation as a function of the chord distance along the preferred axis
for sizes $L = 40, 60, 80$. The solid line indicates the best stretched exponential fit to the $L=80$ data.
No fit of power law form is a plausible match to the data.}
\end{figure}

\begin{figure}
\begin{center}
\ifthenelse {\boolean{PRLVERSION}}
{\includegraphics{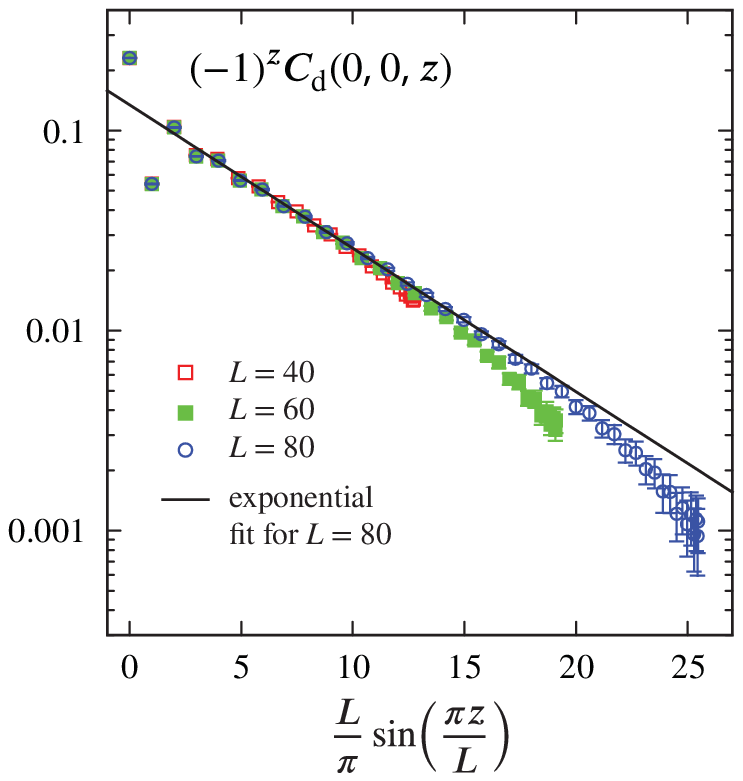}}
{\includegraphics{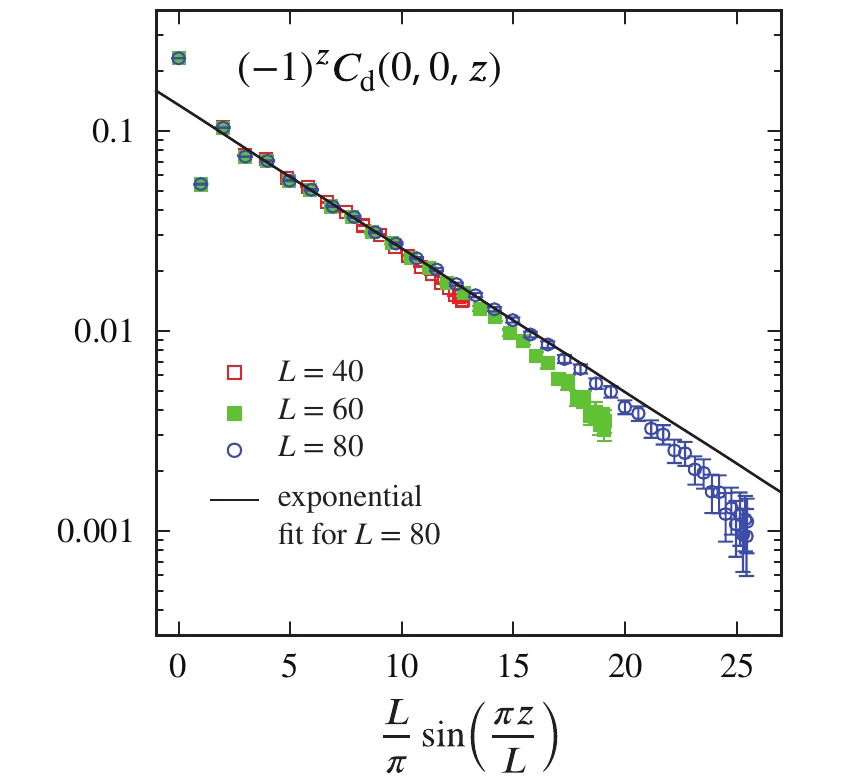}}
\end{center}
\vskip-4mm
\caption{\label{FIG:bond_corr}\co 
The magnitude of the alternating bond correlation as a function of the chord distance along the preferred axis
for sizes $L = 40, 60, 80$. The solid line indicates the best standard exponential fit to the $L=80$ data.}
\end{figure}

As an example, we present in Figs.~\ref{FIG:spin_corr} and \ref{FIG:bond_corr} 
the spin and bond correlation functions, computed at a representative value $a=4.6$. 
In this high anisotropy limit, we find that plotting versus the chord distance $(L/\pi)\sin(\pi z/L)$ substantially 
reduces the finite-size effects. Nonetheless, the liquid state that we observe here is truly
3D, and it is completely different in character from the critical spin liquid that is the 
ground state of the 1D quantum Heisenberg model.

One interesting detail is that although the spin correlation function of this second, easy-axis spin liquid 
falls off much faster than a power law, its decay is not conventionally exponential. Rather, we
find that the function is best fit by a stretched exponential. We determine that its behaviour  in the $L\to\infty$ limit
extrapolates to $(-1)^z C_{\text{s}}(0,0,z) \propto e^{-(z/\ell_{\text{s}})^t}$ 
with a stretching exponent $t = 0.50(4)$ and a decay length $\ell_{\text{s}} = 0.68(5)$.
The bond correlation function, however, shows exponential decay
of the form $(-1)^z C_{\text{b}}(0,0,z) \propto e^{-z/\ell_{\text{b}}}$ 
over a length scale $\ell_{\text{b}} = 7.8(1)$.

{\it Conclusions}.---We have investigated a family of positive-definite, 
short-range RVB wave functions on a layered cubic lattice, where a controllable
anisotropy allows us to interpolate between three points that are effectively 1D ($a=\infty$),
2D ($a=0$), and 3D ($a=1$). 
The phase diagram contains three continuous quantum phase transitions at critical points $a_1 \approx 0.22$,
$a_2 \approx 3.9$, and $a_3 \approx 10$. 

Alternating with the N\'eel antiferromagnet ($a_1 < a < a_2$)
and the valence bond crystal ($a > a_3$) are two phases in which all long-range order is extinguished.
The first is an easy-{\it plane} quantum spin liquid state ($0 < a < a_1$) that has short-range 
spin correlations but critical bond correlations.
The second is an easy-{\it axis} spin liquid ($a_2 < a < a_3$) that is short-ranged with respect to both
spins and bonds.
The existence of this state is surprising, because its wave function is built from exclusively
bipartite valence bonds (only connecting sites in opposite sublattices)
and has a trivial Marshall sign structure~\cite{Marshall55,Beach09a}---a state of affairs that
usually leads to critical bond correlations.

There is some possibility that a cubic 
RVB state of the kind we describe could be realized in ultracold atomic gases in optical lattices~\cite{Bloch08}.
Indeed, proof-of-concept realizations of short-range RVB states on a single, 
four-site plaquette~\cite{Ma11,Nascimbene12} have been acheived. Recently,
even more elaborate short-range bond states have been demonstrated for fermionic cold atoms
in a cubic lattice laser trap with the same anisotropy we consider here~\cite{Grief13}.

We do not have a good intuition for what kind of spin model might have 
this class of wave function as its ground state, but we imagine it must be a highly frustrated one. 
Models on the cubic lattice with competing, {\it nonfrustrating} interactions strong enough to
kill the antiferromagnetism seem invariably to result in crystalline ground states~\cite{Beach07a}.
Note that we make the distinction between frustration at the level of the model and frustration 
at the level of the wavefunction. The latter is simply a statement about a lack of bipartiteness
and the impossibility of choosing bond amplitudes [such as Eq.~\eqref{EQ:h_def}] that are real-valued and nonnegative.

There are some interesting future directions to consider. 
We expect the gapped RVB liquid to be a topological phase. Hence, its 
wave function should exhibit topological entanglement entropy and a degeneracy that depends 
on the genus of the lattice~\cite{Kitaev06,Levin06,Jiang12b}.
We also expect there to be topological invariants under local updates of the valence bond configurations~\cite{Huse03},
and so it should be straightforward to measure properties specific to each topological sector.
It might also be interesting to extend our calculations to product states of SU($N$) singlets~\cite{Beach09b},
in which case the loop fugacity would scale up with $N$. A likely outcome, as $N$ is increased from 2, is that 
the N\'eel ordered region would shrink. It would be worth checking to see if we are ever left with a direct transition between the two liquid states.

We emphasize that our results differ from the work reported in Ref.~\onlinecite{Nahum11}.
Our wave function describes $S = 1/2$ spin degrees of freedom living in three spatial dimensions.

{\it {\Acknowledgements}}.---This work was supported by a Discovery grant from NSERC of Canada. 
Simulations were performed on the computing facilities of WestGrid and on a local cluster generously 
made available to us by John P.\ Davis.


\end{document}